
%
%
%

\input jnl-0.3.tex

\font\titlefont=cmr10 scaled \magstep3
\def\bigtitle{\null\vskip 3pt plus 0.2fill \beginlinemode \doublespace
\raggedcenter \titlefont}

\gdef\journal#1, #2, #3, 1#4#5#6{
{\sl #1~}{\bf #2}, #3 (1#4#5#6)}

\def\3{\sqrt 3}
\def\d{\nabla}
\def\l{\lambda}
\def\a{\alpha}
\def\KN{Kerr-Newman}
\def\RN{Reissner-Nordstr\"om}
\def\KK{Kaluza-Klein}
\def\cO{{\cal O}}
\def\+{{\scriptscriptstyle +}}
\def\-{{\scriptscriptstyle -}}
\def\rmm{ \left(1 - {r_{\-} \over r} \right)}
\def\rpp{ \left(1 - {r_{\+} \over r} \right)}
\def\p{\partial}

\singlespace
\preprintno{UCSBTH-92-11}
\preprintno{hepth@xxx/9203083}
\preprintno{March, 1992}
\doublespace
\bigtitle Rotating Dilaton Black Holes

\author James H. Horne and Gary T. Horowitz

\affil \ucsb
\centerline{jhh@cosmic.physics.ucsb.edu}
\centerline{gary@cosmic.physics.ucsb.edu}

\abstract
It is shown that an arbitrarily small amount of angular momentum can
qualitatively change the properties of extremal charged black holes
coupled to a dilaton.  In addition, the gyromagnetic ratio of these
black holes is computed and an exact rotating black string solution is
presented.

\endtitlepage
\baselineskip=16pt

\subhead{1. Introduction}

All stationary black holes in general relativity coupled to
electromagnetism are described by the \KN\ solutions. These solutions
contain three parameters, the mass $M$, charge $Q$, and angular
momentum $J$, which satisfy the inequality
$  M^2 \ge Q^2 + J^2/M^2.$
(When this inequality is violated, the event horizon disappears and
the spacetime describes a naked singularity.)  It has been known for a
long time that if one minimally couples certain additional matter
fields such as scalars, neutrinos or charged fermions, then the black
hole solutions do not change.  It has recently become clear that if
one couples gravity to more complicated forms of matter, then the
black hole solutions do change. In particular, if one couples to an
SU(2) Yang-Mills theory, new black hole solutions exist~[\cite{Bizon}]
(although they have been shown to be unstable~[\cite{Wald}]).  If one
includes a Higgs field as well as the gauge field, there are
additional solutions which can be interpreted as black holes inside
nonabelian monopoles~[\cite{Lee}].

In this paper we will focus on perhaps the simplest extension of the standard
matter --- gravity coupled to a Maxwell field and a dilaton. The action is
$$ S=\int d^4x \sqrt{-g} \left(-R +2 (\d \Phi )^2 + e^{-2\a \Phi} F^2
                                      \right) \eqno(action) $$
where $\a$ is a free parameter which governs the strength of the
coupling of the dilaton to the Maxwell field\footnote*{In recent
papers, this parameter has been called $a$. We adopt a different
notation here to avoid confusion with the standard angular momentum
parameter for rotating black holes.}. When $\a=0$, the action reduces
to the usual Einstein-Maxwell-scalar theory. When $\a=1$, the
action~\(action) is part of the low energy action of string theory.
However, we will consider this theory for all values of $\a$. Since
changing the sign of $\a$ is equivalent to changing the sign of
$\Phi$, it is sufficient to consider only $\a \ge 0$.

Nonrotating charged black hole solutions of~\(action) have been
found~[\cite{Maeda},\cite{Garfinkle}] and are reviewed below.  Certain
qualitative features of the solutions turn out to be independent of
$\a$. For example, for a given mass, there is always a maximum charge
that can be carried by the black hole. If the charge is less than this
extremal value, there is a regular event horizon. However, other
qualitative features depend crucially on $\a$.  For example, for
$\a\ne 0$ there is no inner horizon.  More importantly, in the
extremal limit, the area $A$ of the event horizon and the surface
gravity $\kappa$ are discontinuous functions of
$\a$.\footnote{$^\dagger$}{In the thermodynamic description, the
entropy is simply $S = {1 \over 4} A$ and the temperature is $T =
{\kappa \over 2 \pi}$.  Since it has been argued that the
thermodynamic interpretation breaks down near the extremal
limit~[\cite{Preskill}], we will focus on the geometrically
well-defined quantities $A$ and $\kappa$.} This has led to the
view~[\cite{Holzhey}] that properties of black holes are not
universal, but depend on the details of the matter lagrangian.
In~[\cite{Holzhey}], it was shown that for $\a > 1$, infinite
potential barriers form around extremal black holes.  This led to the
interpretation that the extremal black holes behave like elementary
particles.

However, the black hole solutions which have been studied so far
represent only a special case of the general black hole solution since
rotation has been ignored.  To see whether qualitative properties of
generic black holes depend on the matter content, one must consider
rotating charged black holes. This is the subject of the present
investigation.  We have not yet found explicit solutions describing
rotating charged black holes for arbitrary $\a$. However for $\a =
\3$, the solutions are known~[\cite{Frolov}].  By considering this one
exact solution and some general arguments, we are led to the
conclusion that at least in terms of properties such as the area of
the event horizon and the surface gravity, rotating black holes do
{\it not} depend qualitatively on $\a$.  For example, for $\a = \sqrt
3$, the area of the event horizon of an extremal black hole is simply
proportional to the angular momentum,
$$ A = 8\pi |J| \, .\eqno(angearly)$$
Thus the area is nonzero in general, and vanishes only in the limit of
no rotation. In addition, the surface gravity vanishes in the extremal
limit whenever the angular momentum is nonzero.  These properties are
more analogous to the usual \KN\ solution ($\a = 0$) than the
nonrotating $\a=\3$ solution. Nevertheless, we will argue that when
the quantum effects of particle creation are taken into account, the
late time behavior of rotating dilaton black holes still resembles
that of elementary particles rather than traditional black
holes.

Although we do not yet have explicit rotating black hole
solutions for all values of
$\a$, we will present the solutions in the limit of slow rotation. This
is sufficient to compute the gyromagnetic ratio of the black hole
\ie the ratio of the magnetic dipole moment to the angular momentum.
This is of interest since in the Einstein-Maxwell theory, black holes
have a gyromagnetic ratio corresponding to $g=2$ (the value for
electrons) rather than $g=1$ (the value for classical matter). We will
see that this is modified when $\a \ne 0 $.  Additionally, extended
black hole or black string solutions to low energy string theory have
recently been found in various
dimensions~[\cite{Horowitz},\cite{HH},\cite{HHS}]. We will present an
exact rotating charged black string solution in five dimensions which
is closely related to the rotating dilaton black hole with $\a = \3$.

We begin by reviewing the unrotating charged black hole solutions
to~\(action).  The field equations are
$$   \d_\mu (e^{-2\a\Phi}F^{\mu\nu}) = 0  \eqno(fuveom)$$
$$   \d^2 {\Phi}+ {\a \over 2} e^{-2\a\Phi}F^2 = 0   \eqno(dileom)$$
$$    R_{\mu\nu} = 2 \d_\mu \Phi  \d_\nu \Phi +2 e^{-2\a\Phi} F_{\mu\rho}
      {F_\nu}^\rho
	-\half g_{\mu\nu} e^{-2\a\Phi} F^2 \,. \eqno(meteom) $$
The spherically symmetric solutions take the
form~[\cite{Maeda},\cite{Garfinkle}]
$$    ds^2 = -\l^2 dt^2 + {dr^2 \over \l^2} + R^2 d\Omega  \eqno(metgen)$$
$$    e^{2\Phi}  = \rmm^{2\a \over 1+\a^2}
                                                          \eqno(dilgen)$$
$$    F_{tr} = {Q \over r^2} \,. \eqno(fuvgen) $$
where
$$ \l^2 = \rpp \rmm^{1-\a^2 \over 1+\a^2} \eqno(lambdadef)$$
and
$$  R= r \rmm^{\a^2 \over 1+\a^2} \,. \eqno(rdef) $$
The two free parameters $r_{\+}$, $r_{\-}$  are related to the
physical mass and charge by
$$ M = { r_{\+} \over 2}
     + \left({1-\a^2 \over 1+\a^2} \right) {r_{\-}\over 2} \eqno(massrprm)$$
$$ Q =  \left({r_{\+} r_{\-} \over 1+\a^2}\right)^\half \>. \eqno(qrprm)$$
When $\a=0$ this solution reduces to the standard \RN\ solution of
Einstein-Maxwell theory. However, for $ \a\ne 0$ the solution is
qualitatively different. For all $\a$, the surface $r=r_{\+}$ is an
event horizon. The surface $r=r_{\-}$ is a curvature singularity
except for the case $\a=0$ when it is a nonsingular inner
horizon\footnote*{This is consistent with the idea that the inner
horizon is unstable in the Einstein-Maxwell theory.}. Thus they
describe black holes only when $r_{\-} < r_{\+}$. In the extremal
limit, $r_{\-} = r_{\+}$, it is clear from~\(metgen) and~\(rdef) that
for $\a\ne 0$ the area of the event horizon goes to zero.  The surface
gravity $\kappa$ is
$$  \kappa = {1\over 2 r_{\+}} \left(1-{r_{\-}
              \over r_{\+}}\right)^{1-\a^2 \over 1+\a^2} \>. \eqno(surface)$$
Thus, for $\a <1$ the surface gravity goes to zero in the extremal
limit, for $\a=1$ it approaches a constant, and for $\a>1$ it
diverges. Holzhey and Wilczek~[\cite{Holzhey}] have shown that for $\a
>1$ as the black hole approaches its extremal limit, there are
potential barriers outside the horizon which increase without bound.
Thus nearly extremal black holes do not absorb low energy incoming
radiation. In addition, any radiation which is absorbed by the black
hole is likely to be radiated very quickly since the temperature is
proportional to $\kappa$. The nearly extremal black holes thus act
more like elementary particles than traditional black holes.

\subhead{2. Rotating Black Holes}

Let us now consider rotating black holes. When $\a = 0$, the
rotating charged solution is the familiar \KN\ solution
$$\eqalign{
ds^2 = & - \left( {\Delta - a^2 \sin^2\theta \over \Sigma} \right) \, dt^2
   - { 2 a \sin^2 \theta (r^2 + a^2 - \Delta) \over \Sigma} \,dt \, d\phi \cr
   & + \left[ {(r^2 + a^2)^2 - \Delta a^2 \sin^2 \theta \over \Sigma} \right]
                                        \sin^2 \theta \, d\phi^2
       + { \Sigma \over \Delta} \, dr^2 + \Sigma \, d\theta^2 \> , \cr
A_{t} = &{Q r \over \Sigma} \>, \quad
A_{\phi} = -{ a Q r \sin^2 \theta \over \Sigma} \> , \quad
\Phi = 0 \>,}   \eqno(kerrnewman)  $$
where
$$ \Sigma = r^2 + a^2 \cos^2 \theta \>, \eqno(sigmadef)$$
$$ \Delta = r^2 + a^2 + Q^2 - 2 M r \>, \eqno(deltadef)$$
and $a$ is the angular momentum $J$ divided by the mass. This solution has
both an event horizon and an inner horizon at the zeros of $\Delta$.
Perhaps the key difference between a rotating and
nonrotating black hole is the existence of an ergosphere. (For
a general discussion, see~[\cite{Waldbook}]). In the
\KN\ solution, $g_{tt} >0$ in a region outside the event horizon. This allows
energy to be extracted classically either by the Penrose process for point
particles or by superradiant scattering for fields. However, the area of the
event horizon cannot decrease in these classical processes.
The area
can be expressed as
$$ A = 8\pi \left[\sqrt{D^2 + Q^4/4} + \sqrt{D^2 -J^2}\right] \eqno(knarea)$$
where $D^2 \equiv M^2(M^2-Q^2)$.
The extraction of energy also
reduces the angular momentum in such a way that $A$ increases.

For any $\a$, the rotating uncharged black hole is simply the $Q=0$
limit of \(kerrnewman) and is called the Kerr solution.  The Kerr
metric has an event horizon at $r= M + \sqrt{M^2 - a^2}$ and an inner
horizon at $r= M - \sqrt{M^2 - a^2}$.  The extremal limit has finite
area and zero surface gravity. We have seen that for $\a > 1 $ the
extremal limit of the unrotating charged black hole has zero area and
infinite surface gravity.  Which of these two situations, zero
rotation or zero charge, is likely to have qualitative properties
similar to the general rotating charged black hole? First, consider a
large uncharged black hole which is rotating near its extremal rate.
Suppose we add a small amount of charge. Since the event horizon is
large and stable, we do not expect the charge to drastically change
the geometry in its vicinity.  This suggests that extremal rotating
black holes with small charge will resemble the uncharged black holes.
In particular, the area of the event horizon should be large and the
surface gravity should be small. Furthermore, the potential barriers
outside the horizon should remain modest.  Now, suppose we start with
a nearly extremal unrotating black hole and try to add a small amount
of angular momentum.  In this case, the event horizon is very small
and the potential barriers are very large.  Thus one cannot add
angular momentum without sending it in with a large amount of energy
which can significantly alter the properties of the solution.  Thus
there is little reason to expect that black holes with small angular
momentum will resemble the zero angular momentum limit.

\subhead{3. The \KK\ Solution}

To determine the behavior of extremal black holes with small angular momentum,
as well as to verify the above arguments for large angular momentum, one needs
to examine exact solutions.
There is one value of $\a$ for which an exact solution is known:
$\a = \sqrt 3$~[\cite{Frolov}].
This is well within the $\a > 1$ regime of~[\cite{Holzhey}] where
the extremal black holes are supposed to behave like elementary particles.
For this value of $\a$, the action~\(action) is simply the \KK\ action
which is obtained by dimensionally reducing the
five dimensional vacuum Einstein action.  In other words, given a five
dimensional spacetime with a translational symmetry in a spacelike
direction, define a four dimensional metric $g_{\mu\nu}$, vector potential
$A_\mu$ and dilaton $\Phi$ by
$$ ds^2 = e^{4\Phi/\3} (dx^5 + 2 A_\mu dx^\mu)^2
         + e^{-2\Phi/\3} g_{\mu\nu} dx^\mu dx^\nu \>. \eqno(kakfive)$$
Then the field equations~\(fuveom), \(dileom), and~\(meteom)
are equivalent to the five dimensional vacuum
Einstein equation.

As a result, it is straightforward to find exact solutions for $\a = \3$.
One starts with a four dimensional vacuum solution of Einstein's equation,
takes the product with ${\bf R}$ to obtain a five dimensional translationally
invariant solution, and finally boosts the solution in the extra direction.
When reinterpreted in four dimensions, the solution has nonzero charge and
a nontrivial dilaton field~[\cite{Wiltshire}].
Starting with the Schwarzschild solution, this
procedure yields the charged black hole~\(metgen) for
$\a =\3$.
To obtain rotating black holes, one simply starts with the Kerr solution
and repeats the same procedure\footnote*{Starting
instead with the charged but massless \KN\ solution leads to a
massive \KK\ solution with a naked singularity~[\cite{Clement}].}.
The resulting metric is~[\cite{Frolov}]
$$ \eqalign{
ds^2 = & - { 1 - Z \over B} \, dt^2
   - {2 aZ \sin^2 \theta  \over B \sqrt{1 - v^2}} \, dt \, d\phi\cr
 & + \left[ B (r^2 + a^2) + a^2\sin^2\theta {Z \over B} \right]\sin^2 \theta \,
                                                            d\phi^2
     + B { \Sigma \over \Delta_0} \, dr^2 + B \Sigma \, d\theta^2 \> , }
                                              \eqno(kakdef) $$
where
$$B = \sqrt{ 1 + { v^2 Z \over 1 - v^2}} \, ,
\quad Z = { 2 m r \over \Sigma} \, ,
\quad \Delta_0 = r^2 + a^2 - 2 m r \eqno(kakdefb)$$
and $m$, $a$ are the mass and rotation parameters of the original Kerr
solution
and $v$ is the velocity of the boost.
$\Sigma$ is defined in~\(sigmadef).
The \KK\ solution has a  vector potential
$$A_t = {v \over 2 (1 - v^2)} { Z \over B^2} \, , \quad
A_{\phi} = - a \sin^2 \theta {v \over 2 \sqrt{1 - v^2}} {Z \over B^2} \, ,
                         \eqno(kkpot)$$
and a dilaton field
$$\Phi = - {\sqrt{3} \over 2} \log B \, . \eqno(kkdil)$$
The physical mass $M$, charge $Q$, and
angular momentum $J$ are given by
$$ M = m\left(1+ {v^2 \over 2(1-v^2)}\right) \eqno(kakmass)$$
$$ Q = { mv \over 1-v^2} \eqno(kakcharge)$$
$$ J = {ma \over \sqrt{1-v^2}} \,.\eqno(kakang)$$
It is easy to see that if $v=0$, the solution reduces to the original Kerr
solution. Furthermore, if $a=0$, the solution agrees with~\(metgen)--\(rdef)
if one sets $\alpha = \3$, identifies
$$ \eqalign{ r_\+ = & {2m \over 1-v^2} \cr
             r_\- = & {2m v^2 \over 1-v^2} \cr} \eqno(kakrprm)$$
and does a simple coordinate transformation $r \rightarrow r + r_\-$
in~\(metgen)--\(rdef).
Notice that the extremal limit $r_\- \rightarrow r_\+$ of the unrotating
black hole corresponds, from the five dimensional viewpoint, to boosting
Schwarzschild cross ${\bf R}$ to the speed of light $v=1$ while taking the
unboosted mass $m$ to zero so that the limit is well defined. In particular,
the infinite potential barriers that arise outside the horizon can
be viewed as a result of boosting the finite barriers outside
Schwarzschild.

The first thing to note about the rotating dilaton black hole~\(kakdef)
is that
the causal structure is very similar to the Kerr solution. The metric
appears singular at the zeros of $\Sigma$ and $\Delta_0$. The first
turns out to be a true curvature singularity at $r=0$, $\theta = \pi/2$
(the ring singularity) and the second is a coordinate effect associated
with regular horizons. There is  an event horizon at
$r = m + \sqrt{m^2 - a^2} $ and a {\it nonsingular} inner horizon at
$r = m - \sqrt{m^2 - a^2} $. Thus  in the presense of rotation, the
dilaton does not destroy the inner horizon.
The extremal limit corresponds to $m^2=a^2$. (Notice that this condition is
independent of $v$ --- an extremal black hole
remains extremal under the boost.)
The extremal limit can be reexpressed as $J^2 = C^2$
where $C\equiv m^2/\sqrt{1-v^2}$ depends on $M$ and $Q$,
but is independent of the
angular momentum $J$. It is easy to show that
for a general black hole, the area of the event horizon is
$$ A = 8\pi \left[ C + \sqrt{C^2 - J^2}\right] \,. \eqno(kakarea)$$
This shows that in the extremal limit, the  area is simply proportional to
the angular momentum:
$ A = 8\pi |J|$.
This clearly demonstrates how the earlier result that the area vanishes
for nonrotating black holes
is modified by rotation. Although the area is  nonzero in general, it
clearly has a different dependence on the physical parameters $M, Q, J$
than the \KN\ solution~\(knarea).

Next we consider the surface gravity of the black hole. This
is~[\cite{Frolov}]
$$ \kappa = {\sqrt{(1-v^2)(m^2 - a^2)} \over 2 m [ m+ \sqrt{m^2 - a^2}]}
                                                      \, .\eqno(kaktemp)$$
If the angular momentum is nonzero, it follows immediately from~\(kaktemp)
that in the extremal
limit the surface gravity goes to zero. For the nonrotating black hole,
the surface gravity reduces to $\kappa=\sqrt{1-v^2}/4 m$ which agrees
with~\(surface)
after using eq.~\(kakrprm)
and setting $\a = \sqrt 3$. In this case the surface gravity
clearly diverges in the extremal limit.
Thus one can view the surface gravity of the extremal black hole
(somewhat heuristically) as being given by $\kappa = \delta (J)$.

Finally we consider the angular velocity of the horizon $\Omega$.
This is defined by the condition that the Killing vector $(\p /\p t)
+ \Omega (\p /\p \phi)$ is null at the event horizon. For the
\KK\ black hole~\(kakdef), the angular velocity is~[\cite{Frolov}]
$$  \Omega = {a\sqrt{(1-v^2)}\over 2 m [ m+ \sqrt{m^2 - a^2}]} \,.
                                              \eqno(kakom) $$
This also exhibits an interesting discontinuity. If one considers nonextremal
black holes and takes the limit as the angular momentum goes to zero,
then $\Omega$ goes to zero as expected. However, if one considers extremal
black holes and takes the limit as $a$ goes to zero (keeping the
physical mass $M$ fixed) one finds that $\Omega$ diverges. This is because
the horizon is shrinking to zero size as the angular momentum
decreases.

We now briefly consider the question of what is the endpoint of
Hawking evaporation if one starts with a nonextremal black hole in
this theory. Let us first consider a nonrotating black hole with
initial mass $M$ and charge $Q$. Since the lagrangian~\(action) does
not include any charged particles, it is clear that the charge cannot
be radiated away. The mass will decrease toward the extremal value and
the temperature will increase. However the potential barriers outside
the black hole will also increase. It seems likely that the black hole
only asymptotically approaches its extremal limit. It is amusing to
consider this from the standpoint of the five dimensional theory.  The
five dimensional theory is more general than~\(action) because modes
depending on the extra dimension correspond to massive charged fields
in the four dimensional theory.  As we have seen, a charged black hole
is simply a boosted form of Schwarzschild cross ${\bf R}$. If the
radius of the compact direction is much smaller than the wavelength of
the Hawking radiation, then the momentum in this direction cannot
change.  (This is the five dimensional analog of the fact that the
charge is constant.)  However the mass will decrease. In order to
conserve momentum it must speed up.  The temperature increases and
hence the wavelength of the emitted radiation decreases.  This process
continues until the wavelength is of order the size of the compact
direction at which point the momentum can change. However, at this
point the potential barriers are so large that it is unlikely that the
black hole will be able to completely radiate away its momentum.

Now we include angular momentum. There are two basic facts about the
quantum mechanics of rotating black holes. The first is that Hawking
radiation carries away angular momentum. The ratio of angular momentum
to energy carried away depends on the spin of the matter field and
increases with increasing spin~[\cite{Page}].  The second fact is that
rotating black holes are never quantum mechanically stable.  Even at
zero temperature there is radiation due to the ergosphere. This
spontaneous emission is closely related to the classical
superradiance. It occurs only for modes $e^{- i \omega t} e^{in\phi}$
satisfying $ \omega /n < \Omega$ \ie when the angular velocity of the
mode is less than the angular velocity of the black hole. A rough
estimate of the rate of mass loss due to radiation from the ergosphere
is~[\cite{Unruh}]
$$    {d M \over dt} \sim \Omega^2 \,. \eqno(ergorad)$$
Thus a rotating dilaton black hole will  radiate away most of its
angular momentum. The area of its event horizon will decrease and the potential
barriers outside the horizon will grow.

Does a rotating dilaton black hole, at late times, resemble an
elementary particle? If the black hole radiates away all its angular
momentum before it approaches the extremal limit, then its evolution
from that point will be the same as in the nonrotating case. If
instead the black hole reaches the extremal limit before radiating
away its angular momentum, then its thermal radiation~\(kaktemp) will
vanish, but the radiation from the ergosphere~\(ergorad) will become
important. As more angular momentum is radiated away, the ergosphere
radiation will increase, and so will the potential barriers outside
the black hole. Thus the late stages of this evolution will resemble
the situation described in~[\cite{Holzhey}]. Extremely high potential
barriers surround a black hole which is radiating rapidly inside the
barriers. In the rotating case the radiation is due to the ergosphere
and in the nonrotating case the radiation is thermal.  Thus an
extremal black hole should behave in a similar fashion in particle
scattering whether it has a small amount of angular momentum or not.
However, one must keep in mind that in both cases the semi-classical
description breaks down in the late stages of evaporation when the
horizon becomes very small.

\subhead{4. Slowly Rotating Black Holes}

It is straightforward to solve the
equations~\(fuveom)--\(meteom) to first order in the angular momentum
parameter $a$. This is because most of the metric
components depend only on $a^2$. Unfortunately, many of the
interesting physical quantities also depend only on $a^2$, but we can
still extract some useful information from the first order solution.

Let us begin with the unrotating black hole for arbitrary $\a$,
eq.~\(metgen), and consider the effect of adding a small amount of rotation
$a$ to the black hole. We will discard any terms involving $a^2$ or
higher powers in $a$. Inspection of the \KN\ and \KK\ solutions
shows that the only term in the metric that changes to
$\cO(a)$ is $g_{t\phi}$. Similarly, the dilaton does not change to
$\cO(a)$, and $A_{\phi}$ is the only component of the vector potential that
changes to $\cO(a)$. However, to this order, the change in $A_{\phi}$
is uniquely determined by the charge.

Thus, for arbitrary $\a$, we can assume the following form of the metric
$$ \eqalign{ ds^2 = & - \rpp \rmm^{ 1 - \a^2 \over 1 + \a^2} dt^2
                      + { dr^2 \over \rpp \rmm^{ 1- \a^2 \over 1 + \a^2}}
                      + r^2 \rmm^{ 2 \a^2 \over 1 + \a^2} d\Omega \cr
                    & - 2 a f(r) \sin^2\theta \,dt\,d\phi, \cr}\eqno(first)$$
where $f(r)$ is a function to be determined.
The dilaton and potential should be
$$ \Phi = {\a \over 1 + \a^2} \log \rmm, \quad A_t = {Q \over r}, \quad
 A_{\phi} = - a \sin^2\theta {Q \over r}. \eqno(firstdil)$$
Using the linearized form of the equations of motion~\(fuveom)--\(meteom),
we see that eqs.~\(first)
and~\(firstdil) are solutions as long as
$$ \eqalign{ f(r) = & { r^2 (1 + \a^2)^2 \rmm^{2 \a^2 \over 1 + \a^2} \over
             (1 - \a^2)(1 - 3 \a^2) r_{\-}^2 } \cr
          & -  \rmm^{ 1 - \a^2 \over 1 + \a^2}
          \left( 1
               + { (1 + \a^2)^2 r^2 \over (1 - \a^2)(1 - 3 \a^2) r_{\-}^2}
               + { (1 + \a^2) r \over (1 - \a^2) r_{\-}}
               - { r_{\+} \over r}
          \right). \cr} \eqno(firstsol) $$
In deriving $f(r)$, we have fixed an integration constant arising
from the equations of motion so that $f(r) \rightarrow c/r$ as
$r \rightarrow \infty$. This is the desired asymptotic behavior,
and completely determines $f(r)$.

It is straightforward to verify that $f(r)$ agrees with the \KN\ solution
when $\a=0$, and with the \KK\ solution~\(kakdef) when $\a = \sqrt{3}$ (after
using~\(kakrprm), shifting $r \rightarrow r + r_{\-}$ and
rescaling $a$ by a constant).
At first glance, $f(r)$ appears to diverge in the string limit,
$\a=1$. This is not actually the case; when $\a = 1$,
$$f(r)_{\a=1} =  {2 r^2 \over r_{\-}^2} \rmm \log \rmm
               - 1 + {2 r \over r_{\-}} + {r_{\+} \over  r}.
                                     \eqno(stringf)$$
Similarly, $\a = 1/\sqrt{3}$ appears singular in the general
solution~\(firstsol), but is also well-behaved. Thus,
$f(r)$ as given in eq.~\(firstsol) is the correct solution
to $\cO(a)$ for all $\a$.

We argued earlier that for $\a >1$,
a small amount of rotation should produce a large
change in the geometry close to the horizon of a nearly extremal
black hole.
We can  see  this explicitly from~\(firstsol).
In the extremal limit, as $r$ approaches $r_{\+} = r_{\-}$,
the second term in $f(r)$ diverges for $\a > 1$. Thus adding even
a little rotation drastically changes the unrotating solution near the horizon.

What else can we learn from the slowly rotating black hole?  The
surface gravity and area of the event horizon do not change to
$\cO(a)$. However, the angular momentum $J$ and the magnetic dipole
moment $\mu$ first appear at this order. The angular momentum is
related to the asymptotic form of the metric by
$$g_{t\phi} {\buildrel  {r \rightarrow \infty} \over \longrightarrow}
            - {2 J \over r} \sin^2\theta + \cO({1 \over r^2}), \eqno(angdef)$$
which gives
$$J = {a \over 2}
 \left( r_{\+} + {3 - \a^2 \over 3 (1 + \a^2)} r_{\-}\right). \eqno(firstang)$$
The magnetic moment is related to the asymptotic form of the vector potential
by
$$A_{\phi} {\buildrel  {r \rightarrow \infty} \over \longrightarrow}
            - {\mu \sin^2 \theta \over r} + \cO({1 \over r^2}), \eqno(magdef)$$
so that
$$\mu = a Q .\eqno(firstmag)$$
If we  define  a parameter $g$ in the usual way by
$$    \mu = g {QJ\over 2M} \, ,\eqno(magmomdefa)$$
then we find
$$g = 2 - { 4 \a^2 r_{\-} \over (3 - \a^2) r_{\-} + 3 (1 + \a^2) r_{\+}}.
                                    \eqno(gyro)$$
When $\a =0$, we recover the well known, but still remarkable fact
that the \KN\ black hole
has $g=2$ just like the electron (up to quantum corrections).
For $\a \ne 0$, $g$ still
approaches 2 in the limit
of small charge. However, as the charge increases, $g$ decreases to a minimum
of  $g = 2 - 2\a^2 / (\a^2 +3)$ for the extremal black hole. So for
string theory with $\a^2 = 1$, ${3 \over 2} \le g \le 2$ and
for \KK\ theory with $\a^2 = 3$, $1 \le g \le 2$.

\subhead{5. Rotating Black Strings}

The \KK\ black hole was constructed by taking a four dimensional uncharged
black hole,
adding an extra dimension, Lorentz boosting, and then using
a \KK\ reduction~[\cite{Frolov}]. This procedure is very similar to
the procedure  recently used to generate axion charged
black string solutions~[\cite{HHS}]. The only difference is that instead of
using a \KK\ reduction in the last step, one uses a
$\sigma$-model
duality transformation~[\cite{Buscher}].
This transformation relates solutions to the low energy action of
string theory: eq.~\(action) with $\a =1$ plus a term
involving the three form $H_{\mu\nu\rho}$ (which is the curl of a two form
$B_{\mu\nu}$). Sigma model duality is most conveniently described in terms
of the metric that the string directly couples to. This is equal to
the metric  appearing in \(action) multiplied by
a  (dimension dependent) power of $e^{\Phi}$. In terms of the string
metric, the (five dimensional) low energy string action is
$$S = \int d^5 x \sqrt{-g} e^{-2 \Phi} \left( -R - 4 (\d \Phi)^2 +
               {1 \over 12} H^2 \right) \,, \eqno(saction)$$
where we have set the Maxwell field to zero.

To obtain rotating black string solutions to \(saction),
begin with the Kerr black hole. Since $\Phi = 0$,  this
is a solution to the string equations of motion.
Add on an extra flat direction $x$, and
Lorentz boost $t \rightarrow (t + v x)/\sqrt{1 - v^2}$,
$x \rightarrow (x + v t)/\sqrt{1 - v^2}$. Now
using the $\sigma$-model duality transformation,
we obtain a five dimensional rotating, charged black string
with metric
$$\eqalign{ ds^2 = & - { 1 - Z \over B^2} \, dt^2
       - { 2 a Z \sin^2 \theta \over B^2 \sqrt{1 - v^2}} \, dt \, d\phi
       + \left[ (r^2 + a^2) + a^2 \sin^2 \theta {Z \over B^2} \right]
                                            \sin^2 \theta\, d\phi^2  \cr
       & + { \Sigma \over \Delta_0} \, dr^2
       + \Sigma \, d\theta^2
         + B^{-2} \, dx^2     \cr
                                 \cr}  \eqno(rotmet)$$
where $\Sigma$, $\Delta_0$, $Z$, and $B$ were defined previously,
with a dilaton field
$$\Phi = - \log B \eqno(rotdil)$$
and antisymmetric tensor field
$$B_{xt} = {v \over (1 - v^2)} { Z \over B^2} \, , \quad
B_{x\phi} = - a \sin^2 \theta {v \over \sqrt{1 - v^2}} {Z \over B^2} \, .
                         \eqno(rotpot)$$
It is straightforward to use the higher dimensional generalizations of
the Kerr solution~[\cite{Myers}] to derive rotating black strings in
higher dimensions.

The properties of the rotating black string~\(rotmet) are very similar
to the Kaluza-Klein black hole.  Indeed, if $d\tilde s^2$ denotes the
\KK\ metric~\(kakdef), then the black string metric~\(rotmet) is
simply $ds^2 = e^{\Phi} d\tilde s^2 + e^{2 \Phi} dx^2$.  The mass per
unit length, axion charge per unit length and angular momentum per
unit length are given by~\(kakmass), \(kakcharge) and~\(kakang).  The
rotating black string has an event horizon and nonsingular inner
horizon at
$r = m \pm \sqrt{ m^2 - a^2}$, and a curvature singularity
at $r=0$, $\theta = \pi/2$. The surface gravity of
the black string is equal to that of the \KK\ solution~\(kaktemp).

The extremal limit of the rotating black string is quite different
from the extremal limit of the unrotating black string~[\cite{HHS}].
In the latter case, the extremal limit corresponds to the field
outside a straight fundamental string~[\cite{Dabholkar}], and is boost
invariant in the $x$-$t$ plane. The extremal limit of the rotating
black string is not boost invariant, and it is not clear whether it
can be interpreted as the field outside of an excited fundamental
string.

\subhead{Acknowledgements}
We would like to thank D.~Wiltshire for
bringing~[\cite{Frolov}],~[\cite{Wiltshire}], and~[\cite{Clement}]
to our attention. This work was supported in
part by NSF Grant PHY-9008502.

\references

\baselineskip=16pt

\refis{Buscher} T.~Buscher,
\journal Phys. Lett., B201, 466, 1988;
\journal Phys. Lett., B194, 59, 1987.

\refis{Holzhey} C.~Holzhey and F.~Wilczek,
``Black Holes as Elementary Particles,''
IAS preprint IASSNS-HEP-91/71, December 1991.

\refis{HH} J.~Horne and G.~Horowitz,
\journal Nucl. Phys., B368, 444, 1992.

\refis{HHS} J.~Horne, G.~Horowitz, and A.~Steif,
\journal Phys. Rev. Lett., 68, 568, 1992.

\refis{Horowitz} G.~Horowitz and A.~Strominger,
\journal Nucl.~Phys., B360, 197, 1991.

\refis{Garfinkle} D.~Garfinkle, G.~Horowitz and A.~Strominger,
\journal Phys.~Rev., D43, 3140, 1991.

\refis{Maeda} G.~Gibbons and K.~Maeda,
\journal Nucl.~Phys., B298, 741, 1988.

\refis{Dabholkar} A.~Dabholkar, G.~Gibbons, J.~Harvey and F.~Ruiz,
\journal Nucl. Phys., B340, 33, 1990.

\refis{Frolov} V.~Frolov, A.~Zelnikov, and U.~Bleyer,
\journal Ann. Phys. (Leipzig), 44, 371, 1987.

\refis{Wiltshire} H.~Leutwyler,
\journal Arch.~Sci., 13, 549, 1960;
P.~Dobiasch and D.~Maison,
\journal Gen. Rel. Grav., 14, 231, 1982;
A.~Chodos and S.~Detweiler,
\journal Gen. Rel. Grav., 14, 879, 1982;
G.~Gibbons and D.~Wiltshire,
\journal Ann. Phys., 167, 201, 1986,
{\it erratum}, {\it ibid.} {\bf 176} (1987) 393.

\refis{Myers} R.~Myers and M.~Perry,
\journal Ann. Phys., 172, 304, 1986.

\refis{Page} D.~Page,
\journal Phys. Rev., D14, 3260, 1976.

\refis{Bizon} P.~Bizon,
\journal Phys. Rev. Lett., 64, 2844, 1990;
M.~Volkov and D.~Gal'tsov,
\journal Sov. J. Nucl. Phys., 51, 1171, 1990;
H.~Kunzle and A.~Masood-ul-Alam,
\journal J. Math. Phys., 31, 928, 1990.

\refis{Wald} P.~Bizon and R.~Wald,
\journal Phys. Lett., B267, 173, 1991.

\refis{Lee} Kimyeong~Lee, V.~Nair and E.~Weinberg,
``Black Holes in Magnetic Monopoles,''
Columbia preprint CU-TP-539, December 1991.

\refis{Preskill} J.~Preskill, P.~Schwarz, A.~Shapere, S.~Trivedi,
and F.~Wilczek,
\journal Mod. Phys. Lett., A6, 2353, 1991.

\refis{Unruh} W.~Unruh,
\journal Phys. Rev., D10, 3194, 1974.

\refis{Waldbook} R.~Wald,
{\it Gereral Relavitity} (U.~of Chicago Press, Chicago), 1984.

\refis{Clement} G.~Cl\'ement,
\journal Gen. Rel. Grav., 18, 137, 1986.

\endreferences

\endit\end